\documentclass[12pt]{iopart}

\usepackage{iopams,graphicx,amssymb}
\eqnobysec

\def\Dm{\widetilde{\cal D}_{\mu}}
\def\D{{\cal D}}

\def\S{{\cal S}}
\def\bfr{{\bf r}}
\def\bfx{{\bf x}}
\def\bfv{{\bf v}}

\begin{document}

\title[Effects of turbulent mixing on critical behaviour in the presence
of compressibility] {Effects of turbulent mixing on critical
behaviour in the presence of compressibility: Renormalization
group analysis of two models}

\author{ N V Antonov and A S Kapustin}

\address{Department of Theoretical Physics, St.~Petersburg State University
\\ Uljanovskaja 1, St.~Petersburg, Petrodvorez, 198504 Russia}

\ead{nikolai.antonov@pobox.spbu.ru}

\begin{abstract}
Critical behaviour of two systems, subjected to the turbulent
mixing, is studied by means of the field theoretic renormalization
group. The first system, described by the equilibrium model {\it
A}, corresponds to relaxational dynamics of a non-conserved order
parameter. The second one is the strongly non-equilibrium
reaction-diffusion system, known as Gribov process and equivalent
to the Reggeon field theory. The turbulent mixing is modelled by
the Kazantsev--Kraichnan ``rapid-change'' ensemble:
time-decorrelated Gaussian velocity field with the power-like
spectrum $\propto k^{-d-\xi}$. Effects of compressibility of the
fluid are studied. It is shown that, depending on the relation
between the exponent $\xi$ and the spatial dimension $d$, the both
systems exhibit four different types of critical behaviour,
associated with four possible fixed points of the renormalization
group equations. Three fixed points correspond to known regimes:
Gaussian fixed point, original model without mixing and passively
advected scalar field. The most interesting fourth point
corresponds to a new type of critical behaviour, in which the
nonlinearity and turbulent mixing are both relevant, and the
critical exponents depend on $d$, $\xi$ and the degree of
compressibility. The critical exponents and regions of stability
for all the regimes are calculated in the leading order of the
double expansion in two parameters $\xi$ and $\varepsilon=4-d$.
For the both models, compressibility enhances the role of the
nonlinear terms in the dynamical equations: the region in the
$\varepsilon$--$\xi$ plane, where the new nontrivial regime is
stable, is getting much wider as the degree of compressibility
increases. For the incompressible fluid, the most realistic values
$d=3$ and $\xi=4/3$ (Kolmogorov turbulence) lie in the region of
stability of the passive scalar regime. If the compressibility
becomes strong enough, the crossover in the critical behaviour
occurs, and these values of $d$ and $\xi$ fall into the region of
stability of the new regime, where the advection and the
nonlinearity are both important. In its turn, turbulent transfer
becomes more efficient due to combined effects of the mixing and
the nonlinear terms.
\end{abstract}


\maketitle

\section{Introduction} \label{sec:Intro}

Numerous systems of very different physical nature exhibit interesting
singular behaviour in the vicinity of their critical points
(second-order or continuous phase transitions). Their correlation functions
reveal self-similar behaviour with rather universal
critical dimensions: they depend only on few global characteristics of
the system (like symmetry or space dimensionality).
Quantitative description of critical behaviour is provided by the
renormalization group (RG). In the RG approach, possible types of critical
regimes (universality classes) are associated with infrared (IR) attractive
fixed points of renormalizable field theoretic models. Most typical
{\it equilibrium} phase
transitions belong to the universality class of the $O_{n}$-symmetric
$\psi^{4}$ model of an $n$-component scalar order parameter.
Universal characteristics of the critical behaviour depend only on $n$ and
the space dimensionality $d$ and can be calculated in the form of the
expansion in $\varepsilon=4-d$ or within other systematic perturbation
schemes; see the monographs \cite{Zinn,Book3} and the literature cited
therein. Equilibrium {\it dynamical} critical behaviour is much richer
and less universal; it is described by variety of models; see the
reviews \cite{HH} and chap.~5 of book \cite{Book3}.

Over the past few decades, much attention has been attracted by the
spreading processes and corresponding {\it nonequilibrium} phase
transitions; see e.g.~\cite{Hinr,JT} and the literature cited therein.
Spreading processes are encountered in
physical, chemical, biological, ecological and sociological systems:
autocatalytic reactions, percolation in porous media, forest fires,
epidemic diseases and so on. Depending on the conditions, the spreading
of an ``agent'' (e.g. infectious disease) can either
continue over the whole population, or terminate after some time. In the
first case, the system evolves to a stationary (but {\it not} thermally
equilibrium) active state, in which the density of the agent remains a
fluctuating quantity. In the second case, the system is ``trapped'' in an
absorbing (inactive) state where all the fluctuations cease completely.
Transition between these phases are continuous; they are especially
interesting as examples of nonequilibrium critical behaviour
\cite{Hinr}. The most popular model that describes the spreading
of an agent (e.g., infectious disease) in a reaction-diffusion system
belongs to the universality class of the directed bond percolation
process, also known as simple epidemic process or Gribov process
\cite{Hinr}. In the field theoretic formulation that model is equivalent
to the Reggeon field theory and was also studied within the
RG approach and the $\varepsilon$ expansion  \cite{JT}.

However, it has long been realized that the behaviour of a real system
near its critical point is extremely sensitive to external disturbances,
geometry of the experimental setup, gravity, presence of impurities and
so on; see e.g. the monograph \cite{Ivanov} for the general discussion and
the references. ``Ideal''  critical behaviour of an infinite stationary
system can be obscured by
finite-size effects, finite time of evolution (ageing) and so on.
What is more, some disturbances (randomly distributed impurities or
turbulent mixing) can produce completely new types of critical
behaviour with rich and rather exotic properties, like e.g. expansion in
$\sqrt{\varepsilon}$ rather than in $\varepsilon$  \cite{quench,Satten}.

Those issues become especially important for the nonequilibrium phase
transitions, because the ideal conditions of a ``pure'' stationary critical
state can hardly be achieved in real chemical or biological systems, and
the effects of various disturbances can never be completely excluded.
In particular, intrinsic turbulence effects cannot be avoided in
chemical catalytic reactions or forest fires. One can also speculate that
atmospheric turbulence can play important role for the spreading of an
infectious disease by flying insects or birds.

Investigation of the effects of various kinds of deterministic or chaotic
flows (laminar shear flows, turbulent convection and so on) on the behaviour
of the critical fluids (like binary liquid mixtures near their consolution
points) has shown that the flow can destroy the usual critical behaviour,
typical of the $\psi^{4}$ or Reggeon models. It can change to the mean-field
behaviour or, under some conditions, to a more complex behaviour described
by new nonequilibrium universality classes \cite{Satten}--\cite{AIM}.

In this paper we study effects of turbulent mixing on the critical behaviour
of two systems near their critical points, paying special attention to
{\it compressibility} of the fluid. To describe the phase transitions,
we use two
paradigmatic models of dynamical critical behaviour. The first one is the
equilibrium model {\it A}, which describes purely relaxational dynamics of
a non-conserved order parameter (e.g. in a nematic liquid crystal);
see the reviews \cite{HH} and chap.~5 of book \cite{Book3}. The second one,
referred to as Gribov model in what follows, is the aforementioned model
of a nonequilibrium reaction-diffusion system \cite{Hinr,JT}.

To model the turbulent mixing, we employ the famous Kazantsev--Kraichnan
``rapid-change'' ensemble: time-decorrelated Gaussian velocity field
with the pair velocity function of the form
$\langle vv\rangle \propto \delta(t-t') \, k^{-d-\xi}$, where
$k$ is the wave number and $0<\xi<2$ is a free parameter with the most
realistic (``Kolmogorov'') value $\xi=4/3$.
This model has attracted enormous attention recently because of the insight
it offers into the origin of intermittency and anomalous scaling in the
fully developed turbulence; see the review paper \cite{FGV} and references
therein. The RG approach to that problem is reviewed in \cite{JphysA}.
In the context of our study it is especially important that the
Kazantsev--Kraichnan ensemble allows one to easily model
{\it compressibility} of the fluid, which appears rather difficult
if the velocity is modelled by dynamical equations; see e.g.
discussion in~\cite{Compress}.

The outline of the paper and the main results are the following.
In section~\ref{sec:QFT} we present the detailed description of the models
and their field theoretic formulation.
In section~\ref{sec:Reno} we analyze the ultraviolet (UV) divergences,
relaying upon the power counting and additional symmetry considerations.
We show that the both models, after proper extension,
are multiplicatively renormalizable. Thus we can derive the RG equations
and introduce the RG functions ($\beta$ functions and anomalous dimensions
$\gamma$) in the standard fashion; see section~\ref{sec:RGE}.

In section~\ref{sec:FPS} we show that, depending on the relation between
the spatial dimension $d$ and the exponent $\xi$ in
the velocity correlator, the both models exhibit four different types of
critical behaviour, associated with four fixed points of the corresponding
RG equations. Three fixed points correspond to known regimes:
ordinary diffusion (Gaussian or free field theory), non-interacting
scalar field passively advected by the flow (the nonlinearity in the order
parameter in the original dynamical equations appears unimportant),
and the original critical behaviour of a model without mixing (that is,
model {\it A} or Gribov model). The most interesting fourth point corresponds
to a new full-scale nonequilibrium universality class, in which both
the nonlinearity and turbulent mixing are relevant. The corresponding
critical exponents can be systematically calculated as double expansions
in two parameters: $\xi$ and $\varepsilon=4-d$.

The practical calculation of the renormalization constants, RG functions,
regions of stability and critical dimensions was accomplished to the
leading order (one-loop approximation); some of the results, however, are
exact (valid to all orders of the double $\varepsilon$--$\xi$ expansion).
In general, the critical exponents and the boundaries
between the regions of stability of critical regimes in the
$\varepsilon$--$\xi$ plane depend on the degree of compressibility.

The results obtained are discussed in section~\ref{sec:Conc}.
The main qualitative conclusion of our analysis is that, for the both
models, compressibility enhances the role of the nonlinear terms in the
dynamical equations. The region in the $\varepsilon$--$\xi$ plane, where
the new nontrivial regime is stable (the corresponding fixed point is
attractive), is getting much wider as the degree of compressibility
increases. For the case of incompressible fluid, the most realistic values
$d=3$ and $\xi=4/3$ lie in the region of stability of the passive scalar
regime, that is, the nonlinear terms are irrelevant for the critical
behaviour. If the compressibility becomes strong enough, the crossover
in the critical behaviour occurs, and these values of $d$ and $\xi$
fall into the region of stability of the new regime, where the advection
and the nonlinearity are both important.

These general considerations are illustrated by the example of a cloud
of particles, randomly walking in a nearly critical turbulent medium.
For a weakly compressible fluid, spreading of such a cloud is
determined solely by the turbulent transfer. If the compressibility becomes
strong enough, the new scaling regime comes into play and the transfer
becomes faster, due to combined effects of the mixing and the nonlinear
terms. The corresponding exponent depends explicitly on the degree of
compressibility and differs from the well-known ``1/2 law'' for ordinary
diffusion and from Richardson's ``4/3 law'' for turbulent transfer.

\section{Description of the models. Field theoretic formulation}
\label{sec:QFT}

In the Langevin formulation the models are defined by stochastic
differential equations for the order parameter
$\psi = \psi(t,{\bf x})$:
\begin{eqnarray}
\partial_{t} \psi = \lambda_{0} \left\{ (-\tau_{0} +
\partial^{2}) \psi - V(\psi) \right\} + \zeta  = 0,
\label{stoh}
\end{eqnarray}
where $\partial_{t}= \partial/ \partial t$, $\partial^{2}$ is the
Laplace operator, $\lambda_{0}>0$ is the kinematic (diffusion)
coefficient and $\tau_{0} \propto (T-T_{c})$ is the deviation of
the temperature (or its analog) from the critical value. The
nonlinearity has the form $V(\psi)=u_{0} \psi^{3}/3!$ for the model {\it A}
and $V(\psi)=g_{0} \psi^{2}/2$ for the Gribov process; $g_{0}$ and
$u_{0}>0$ being the coupling constants. The Gaussian random noise
$\zeta=\zeta(t,{\bf x})$ with the zero mean is specified by the
pair correlation function:
\begin{eqnarray}
\langle \zeta (t,{\bf x})\zeta (t',{\bf x'}) \rangle =
2 \lambda_{0}  \delta(t-t')\delta^{(d)}({\bf x}-{\bf x}')
\label{kor1}
\end{eqnarray}
for the model {\it A} and
\begin{equation}
\langle \zeta (t,{\bf x})\zeta (t',{\bf x'}) \rangle = g_{0}\lambda_{0}\,
\psi(t,{\bf x})\,   \delta(t-t')\delta^{(d)}({\bf x}-{\bf x}')
\label{shum}
\end{equation}
for the Gribov process; $d$ is the dimension of the ${\bf x}$ space. The
factor $\psi$ in front of the correlator (\ref{shum}) guarantees that in the
absorbing state the fluctuations cease entirely, while the factor
$2\lambda_{0}$ in (\ref{kor1}) ensures the correspondence to the static
$\psi^{4}$ model. Here and below, the bare (unrenormalized) parameters are
marked by the subscript ``o.'' Their renormalized analogs (without
the subscript) will appear later on.

The stochastic problems (\ref{stoh}), (\ref{kor1}), (\ref{shum}) can be
reformulated as field theoretic models of the
doubled set of fields $\Phi = \{\psi,\psi^{\dag}\}$ with action functional
\begin{eqnarray}
\S(\psi,\psi^{\dag}) =  \psi^{\dag}
\left(-\partial_{t}+\lambda_{0} \partial^{2}- \lambda_{0}\tau_{0}\right)
\psi + \lambda_{0} (\psi^{\dagger})^2 - u_{0} \psi^{\dag}\psi^3/3!
\label{actionA}
\end{eqnarray}
for the model {\it A} and
\begin{eqnarray}
\S(\psi,\psi^{\dag}) =  \psi^{\dag}
(-\partial_{t}+\lambda_{0} \partial^{2}- \lambda_{0}\tau_{0}) \psi
+ \frac{g_{0}\lambda_{0}}{2} \left\{ (\psi^{\dagger})^2\psi -
\psi^{\dagger}\psi^2  \right\}
\label{actionG}
\end{eqnarray}
for the Gribov model.
Here, $\psi^{\dag}=\psi^{\dag}(t,{\bf x})$ is the auxiliary ``response
field'' and the integrations over the arguments of the fields are implied,
for example
\[  \psi^{\dag}\partial_{t}\psi = \int dt \int d{\bf x}
\psi^{\dag}(t,{\bf x})\partial_{t}\psi(t,{\bf x}). \]
The field theoretic formulation means that the statistical averages
of random quantities in the original stochastic problems can be represented
as functional integrals over the full set of fields with weight
$\exp {\cal S}(\Phi)$, and can therefore be viewed as the Green functions
of the field theoretic models with actions (\ref{actionA}), (\ref{actionG}).
In particular, the linear response function of the problems
(\ref{stoh}), (\ref{kor1}), (\ref{shum}) is given by the Green function
\begin{eqnarray}
G=\langle \psi^{\dag}(t, {\bf x}) \psi(t', {\bf x'} ) \rangle =
\int {\cal D}\psi^{\dag} \int {\cal D} \psi\ \,
\psi^{\dag}(t, {\bf x}) \psi(t', {\bf x'})\, \exp {\cal S}(\psi,\psi^{\dag})
\label{respd}
\end{eqnarray}
of the corresponding field theoretic model.\footnote{The continuous
$\delta$-correlated stochastic noise like $\zeta$ in (\ref{shum}), usually
referred to as {\it multiplicative} random noise, requires additional
careful definition; see e.g.~\cite{vanK}. Consistent description is provided
by the so-called It\^{o} time discretization, which is also assumed here.}

The model (\ref{actionG}) corresponds to the standard Feynman
diagrammatic technique with the only bare propagator
$\langle \psi \psi^{\dag} \rangle_{0}$ and two triple vertices
$\sim (\psi^{\dagger})^2\psi$, $\psi^{\dagger}\psi^2$.
In the time-momentum and frequency-momentum representation the propagator
has the form
\begin{eqnarray}
\langle \psi \psi^{\dag} \rangle_{0}(t,k) =
\theta(t) \exp \left\{ - \lambda_{0}
(k^{2}+\tau_{0}) t \right\}, \nonumber \\
\langle \psi \psi^{\dag} \rangle_{0} (\omega,k) =
\frac{1}{-{\rm i}\omega+\lambda_{0} \left(k^{2}+\tau_{0}\right)}.
\label{lines}
\end{eqnarray}
Here $\theta(\dots)$ is the Heaviside step function, so that the propagator
(\ref{lines}) is retarded.\footnote{The It\^{o} discretization
accepted above means that all the diagrams with self-contracted
$\langle \psi^{\dag} \psi \rangle_{0}$ lines should be discarded.
Vanishing of the propagator $\langle \psi^{\dag} \psi^{\dag} \rangle_{0}$
is the general feature of stochastic dynamical models.}
Then from the analysis of the diagrams one can
check that the Green functions built solely from the field $\psi$
or solely from $\psi^{\dag}$ necessarily contain closed circuits of
retarded propagators (\ref{lines}) and therefore vanish identically.

For the functions $\langle \psi^{\dag} \dots \psi^{\dag} \rangle$ this
fact is a general consequence of the causality, which is valid
for any stochastic model; see e.g. the discussion in \cite{Book3}.
Then vanishing of the functions $\langle \psi \dots \psi \rangle$
in the model (\ref{actionG}) can be viewed as a consequence of the
symmetry with respect to the transformation
\begin{equation}
\psi(t,{\bf x})\to \psi^{\dagger}(-t,-{\bf x}),
\quad \psi^{\dagger}(t,{\bf x})\to \psi(-t,-{\bf x}),
\quad g_{0}\to -g_{0}.
\label{symm}
\end{equation}
Reflection of the constant $g_{0}$ is in fact unimportant because, as can
easily be seen, in the model (\ref{actionG}) with trilinear interaction
terms the actual expansion parameter in the perturbation theory is
$g_{0}^{2}$ rather than $g_{0}$ itself; we shall also denote it as
$u_{0}\equiv g_{0}^{2}$ in what follows.

In addition to (\ref{lines}), diagrammatic technique for the model {\it A}
involves the propagator $\langle \psi \psi \rangle_{0}$ of the form
\begin{eqnarray}
\langle \psi \psi \rangle_{0}(t,k) = \frac{1}{(k^{2}+\tau_{0})}\,
\exp \left\{ - \lambda_{0} (k^{2}+\tau_{0})|t| \right\},
\nonumber \\
\langle \psi \psi \rangle_{0}(\omega,k) = \frac{2\lambda_{0}}
{\omega^{2} + \lambda_{0}^{2} (k^{2}+\tau_{0})^{2}},
\label{lines2}
\end{eqnarray}
while the only vertex has the form $\sim \psi^{\dagger}\psi^{3}$.

The Galilean invariant coupling with the velocity field
${\bf v}= \{ v_{i}(t,\bfx) \}$ for the compressible fluid
($\partial _i v_{i} \ne 0$) can be introduced by the replacement
\begin{eqnarray}
\partial_{t}\psi \to \partial_{t}\psi + a_{0}\, \partial_{i}(v_{i}\psi) +
(a_{0}-1) (v_{i} \partial_{i})\psi =
\nabla_{t} \psi + a_{0}(\partial_{i}v_{i}) \psi
\label{nabla}
\end{eqnarray}
in (\ref{stoh}). Here $\nabla_{t} \equiv \partial_{t} + v_{i} \partial_{i}$
is the Lagrangian derivative, $a_{0}$ is an arbitrary parameter and
$\partial_i = \partial /\partial x_{i}$. For the linear advection-diffusion
equation, the choice $a_{0}=1$ corresponds to the conserved quantity $\psi$
(e.g. density of an impurity) while $a_{0}=0$ is referred to as ``tracer''
(concentration of the impurity or the temperature of the fluid; in this
case, the conserved quantity is $\psi^{\dag}$).
As we shall see below, in the presence of nonlinearities in (\ref{stoh}),
it is necessary to simultaneously include the both terms (\ref{nabla}) in
order to ensure multiplicative renormalizability.

In the real problem, the field ${\bf v}(t,{\bf x})$ satisfies the
Navier--Stokes equation. We will employ the rapid-change model \cite{FGV},
where the velocity obeys a Gaussian distribution with zero mean and
the correlation function
\begin{eqnarray}
\langle v_{i}(t, \bfx) v_{j}(t',{\bf x'})\rangle =  \delta(t-t')\,
D_{ij}(\bfr), \quad \bfr = \bfx-{\bf x'}
\label{white}
\end{eqnarray}
with
\begin{eqnarray}
D_{ij}(\bfr) = D_{0}\, \int_{k>m} \frac{d{\bf k}}{(2\pi)^{d}} \,
\frac{1}{k^{d+\xi}}\, \left\{ P_{ij}({\bf k})+\alpha Q_{ij}({\bf
k}) \right\}\, \exp ({\rm i} {\bf kr} ).
\label{Kraich}
\end{eqnarray}
Here $P_{ij}({\bf k}) = \delta_{ij} - k_i k_j / k^2$ and
$Q_{ij}({\bf k})=k_i k_j/k^2$ are the transverse and the longitudinal
projectors, $k\equiv |{\bf k}|$ is the wave number,
$D_{0}>0$ is an amplitude factor and $\alpha>0$ is an arbitrary
parameter. The case $\alpha=0$ corresponds to the incompressible fluid
($\partial _i v_{i}=0$), while the limit $\alpha \to\infty$ at fixed
$\alpha D_{0}$ corresponds to the purely potential velocity field.
The exponent $0<\xi<2$ is a free parameter which can be viewed as a kind
of H\"{o}lder exponent, which measures ``roughness'' of the velocity field;
the ``Kolmogorov'' value is $\xi=4/3$, while the ``Batchelor'' limit
$\xi\to2$ corresponds to smooth velocity. The cutoff in the
integral (\ref{Kraich}) from below at $k=m$, where $m\equiv 1/{\cal L}$ is
the reciprocal of the integral turbulence scale ${\cal L}$, provides IR
regularization. Its precise form is unimportant; the sharp cutoff is the
simplest choice for the practical calculations.

The action functionals for the full set of fields
$\Phi = \left\{ \psi,\psi^{\dag},{\bf v} \right\}$ become
\begin{eqnarray}
\S_{A}(\Phi) = \S_{C}(\Phi) + \lambda_{0} (\psi^{\dag})^{2}
- u_{0}\lambda_{0} \psi^{\dag} \psi^{3} /3!
\label{ActionA}
\end{eqnarray}
for the model {\it A} and
\begin{eqnarray}
\S_{G}(\Phi) = \S_{C}(\Phi)
+ \frac{g_{0}\lambda_{0}}{2} \left\{ (\psi^{\dagger})^2\psi -
\psi^{\dagger}\psi^2  \right\}
\label{ActionG}
\end{eqnarray}
for the Gribov model, where the common part of the actions has the form
\begin{eqnarray}
\S_{C}(\Phi) =  \psi^{\dag} \left\{
-\nabla_{t} + \lambda_{0}\left( \partial^{2}- \tau_{0}\right)
- a_{0} (\partial_{i}v_{i}) \right\} \psi +  \S(\bfv).
\label{ActionC}
\end{eqnarray}
They are obtained from (\ref{actionA}), (\ref{actionG}), by the replacement
(\ref{nabla}) and adding the term corresponding to the Gaussian averaging
over the field $\bfv$ with the correlator (\ref{white}), (\ref{Kraich}):
\begin{eqnarray}
\S(\bfv) = -\frac{1}{2} \int dt \int d{\bf x} \int d{\bf x'}
v_{i} (t,\bfx) D_{ij}^{-1}(\bfr) v_{j} (t, {\bf x'}),
\label{Sv}
\end{eqnarray}
where
\[ D_{ij}^{-1}(\bfr) \propto D_{0}^{-1} r^{-2d-\xi} \]
is the kernel of the inverse linear operation for the function
$D_{ij}(\bfr)$ in (\ref{Kraich}).

In addition to (\ref{lines}), the Feynman diagrams for the models
(\ref{ActionA})--(\ref{ActionC}) involve the propagator
$\langle vv \rangle_{0}$ specified by the relations (\ref{white}),
(\ref{Kraich}) and the new vertex
\begin{equation}
\psi^{\dag} v_{i} V_{i} \psi \equiv -
\psi^{\dag}\left\{ (v_{i}\partial_{i}) \psi + a_{0}(\partial_{i}v_{i})
\right\} \psi.
\label{Vertex}
\end{equation}
In the diagrams it corresponds to the vertex factor
\begin{equation}
V_{i} = - {\rm i} k_{i} - {\rm i} a_{0} q_{i},
\label{VertexF}
\end{equation}
where $k_{i}$ is the momentum argument of the field $\psi$ and
$q_{i}$ is the momentum of $v_{i}$.

The reflection symmetry $\psi, \psi^{\dag} \to -\psi, -\psi^{\dag}$
of the model
{\it A} is preserved in the full model (\ref{ActionA}), while for
the full model (\ref{ActionG}) the relations (\ref{symm}) should be
augmented by the transformation
\begin{equation}
a_{0} \to (1-a_{0}),
\label{symmG}
\end{equation}
as can directly be seen from (\ref{ActionC}) with the aid of integration
by parts.

For the both full models, the role of the coupling constants (expansion
parameters in the perturbation theory) is played by the three parameters
\begin{equation}
u_{0}  \sim \Lambda^{4-d}, \qquad
w_{0} = D_{0}/\lambda_{0} \sim \Lambda^{\xi},
\qquad w_{0}a_{0} \sim \Lambda^{\xi}.
\label{charges}
\end{equation}
The last relations, following from the dimensionality considerations
(more precisely, see the next section), define the typical UV momentum
scale $\Lambda$. From the relations (\ref{charges}) it follows that the
interactions $\psi^{\dagger}\psi^{3}$, $(\psi^{\dagger})^2\psi$ and
$\psi^{\dagger}\psi^2$ become logarithmic (the corresponding coupling
constant $u_{0}$ becomes dimensionless) at $d=4$.
Thus for the single-charge problems (\ref{actionA}), (\ref{actionG}),
the value $d=d_{c}=4$ is
the upper critical dimension, and the deviation $\varepsilon=4-d_{c}$ plays the
part of the formal expansion parameter in the RG approach: the critical
exponents are nontrivial for $\varepsilon>0$ and are calculated as series in
$\varepsilon$.

The additional interactions $\sim \psi^{\dag} v\partial \psi $
of the full models (\ref{ActionA}), (\ref{ActionG})
become logarithmic at $\xi=0$. The parameter $\xi$ is not
related to the spatial dimension and can be varied independently. However,
for the RG analysis of the full problems it is important that
all the interactions become logarithmic at the same time. Otherwise, one
of them would be weaker than the others from the RG viewpoint and it would
be irrelevant in the leading-order IR behaviour. As a result, some of the
scaling regimes of the full model would be overlooked.
In order to study all possible scaling regimes and the crossovers between
them, we need a genuine three-charge theory, in which all the interactions
are treated on equal footing. Thus we will treat $\varepsilon$ and $\xi$ as
small parameters of the same order, $\varepsilon \propto \xi$.
Instead of the plain $\varepsilon$ expansion in the single-charge models,
the coordinates of the fixed points, critical dimensions
and other quantities will be calculated as double expansions in the
$\varepsilon$--$\xi$ plane around the origin, that is, around the point in which
all the coupling constants in (\ref{charges}) become dimensionless.
Similar situation was encountered earlier in various models of turbulence
and complex critical behaviour, e.g. \cite{AHH,Alexa,AIK,AIM,Sak}.

\section{Canonical dimensions, UV divergences and the renormalization}
\label{sec:Reno}

It is well known that the analysis of UV divergences is based on the analysis
of canonical dimensions (``power counting''); see e.g. \cite{Zinn,Book3}.
Dynamical models of the type (\ref{actionA}), (\ref{actionG}) and
(\ref{ActionA})--(\ref{ActionC}), in contrast to static ones, have
two independent scales: the time scale $T$ and the length scale $L$. Thus
the canonical dimension of any quantity $F$ (a field or a parameter) is
characterized by two numbers, the
frequency dimension $d_{F}^{\omega}$ and the momentum dimension $d_{F}^{k}$,
defined such that $[F] \sim [T]^{-d_{F}^{\omega}} [L]^{-d_{F}^{k}}$. These
dimensions are found from the obvious normalization conditions
\[ d_k^k=-d_{\bf x}^k=1,\ d_k^{\omega} =d_{\bf x}^{\omega }=0,\
d_{\omega }^k=d_t^k=0, \ d_{\omega }^{\omega }=-d_t^{\omega }=1 \]
and from the requirement
that each term of the action functional be dimensionless (with
respect to the momentum and frequency dimensions separately).
Then, based on $d_{F}^{k}$ and $d_{F}^{\omega}$,
one can introduce the total canonical dimension
$d_{F}=d_{F}^{k}+2d_{F}^{\omega}$ (in the free theory,
$\partial_{t}\propto\partial^{2}$), which plays in the theory of
renormalization of dynamical models the same role as
the conventional (momentum) dimension does in static problems;
see Chap.~5 of \cite{Book3}.
The canonical dimensions for the models (\ref{ActionA})--(\ref{ActionC})
are given in table~\ref{table1}, including renormalized parameters (without
subscript ``o''), which will be introduced soon.
The fields of the model {\it A} are marked by the subscript {\it A},
while the fields of the Gribov model are marked by {\it G}. The dimensions
of the parameters $\lambda_{0}$, $\tau_{0}$ {\it etc} are the same for the
both models.

\begin{table}
\caption{Canonical dimensions of the fields and parameters in the
models (\protect\ref{ActionA})--(\protect\ref{ActionC}).}
\label{table1}
\begin{tabular}{ccccccccccc}
\br
$F$ & $\psi_{A}$ & $\psi_{A}^{\dag}$ & $\psi_{G}$, $\psi_{G}^{\dag}$ &
$ {\bf v} $ &   $\lambda_{0}$, $\lambda$ &
$\tau_{0}$, $\tau$ &  $m$, $\mu$, $\Lambda$ & $g_{0}^{2}$, $u_{0}$ &
$w_{0}$ & $u$, $w$, $\alpha$, $a_{0}$, $a$ \\
\br
$d_{F}^{k}$ & $d/2-1$ & $d/2+1$ & $d/2$ & $-1$ & $-2$  & 2
& 1 &  $4-d$ & $\xi$ & 0 \\
\mr
$d_{F}^{\omega }$ & 0 & 0& 0 & 1 & 1 & 0 & 0 &  0 & 0 & 0 \\
\mr
$d_{F}$ & $d/2-1$ & $d/2+1$ &
$d/2$ & 1 & 0 & 2 & 1 &  $4-d$  & $\xi$ & 0 \\
\br
\end{tabular}
\end{table}

As already discussed in the end of the previous section, the full models are
logarithmic (all the coupling constants are simultaneously dimensionless) at
$d=4$ and $\xi=0$. Thus the UV divergences in the Green functions manifest
themselves as poles in $\varepsilon = 4-d$, $\xi$ and, in general, their linear
combinations.

The total canonical dimension of an arbitrary 1-irreducible Green function
$\Gamma = \langle\Phi \cdots \Phi \rangle _{\rm 1-ir}$
is given by the relation \cite{Book3}
\begin{equation}
d_{\Gamma }=d_{\Gamma }^k+2d_{\Gamma }^{\omega }= d+2-N_{\Phi }d_{\Phi},
\label{dGamma}
\end{equation}
where $N_{\Phi}=\{N_{\psi},N_{\psi^{\dag}}, N_{v}\}$ are the numbers of
corresponding fields entering into the function $\Gamma$, and the summation
over all types of the fields is implied. The total dimension $d_{\Gamma}$
in logarithmic theory (that is, at $\varepsilon=\xi=0$) is the formal index of the
UV divergence $\delta_{\Gamma}=d_{\Gamma}|_{\varepsilon=\xi=0}$. Superficial UV
divergences, whose removal requires counterterms, can be present only in
those functions $\Gamma$ for which $\delta_{\Gamma}$ is a non-negative
integer.

From table~\ref{table1} and (\ref{dGamma}) we find
\begin{equation}
\delta_{\Gamma}= 6 - N_{\psi} - 3N_{\psi^{\dag}} - N_{v}
\label{IndeA}
\end{equation}
for the model {\it A} and
\begin{equation}
\delta_{\Gamma}= 6 - 2N_{\psi} - 2N_{\psi^{\dag}} - N_{v}
\label{IndeG}
\end{equation}
for the Gribov model.

In dynamical models, the 1-irreducible diagrams without the fields
$\psi^{\dag}$ vanish, and it is sufficient to consider the functions
with $N_{\psi^{\dag}} \ge 1$. For the Gribov model, the functions built
solely from the fields $\psi^{\dag}$ also vanish due to the symmetry
(\ref{symm}), (\ref{symmG}), and it is sufficient to take $N_{\psi}\ge1$
in (\ref{IndeG}). For the model {\it A}, due to the reflection symmetry,
it is sufficient to consider in (\ref{IndeA}) only the functions with
even $N_{\psi^{\dag}}+N_{\psi}$. With these restrictions, the analysis of
the expressions (\ref{IndeA}), (\ref{IndeG}) shows that in the both
models, superficial UV divergences can be present in the following
1-irreducible functions:
\[ \langle \psi^{\dag} \psi \rangle \quad (\delta=2) \quad
{\rm with\ the\ counterterms} \quad \psi^{\dag}\partial_{t}\psi, \
\psi^{\dag}\partial^{2}\psi, \ \psi^{\dag}\psi, \]
\[ \langle \psi^{\dag} \psi v \rangle \quad (\delta=1) \quad
{\rm with\ the\ counterterms} \quad \psi^{\dag} (v\partial) \psi, \
\psi^{\dag} (\partial v) \psi.  \]
For the model {\it A}, superficial divergence can also be present in
the function
\[ \langle \psi^{\dag} \psi\psi\psi \rangle \quad (\delta=0) \quad
{\rm with\ the\ counterterm} \quad \psi^{\dag} \psi^{3}, \]
while for the Gribov model, divergences can also be present in:
\[ \langle \psi^{\dag} \psi\psi \rangle \quad (\delta=0) \quad
{\rm with\ the\ counterterm} \quad \psi^{\dag} \psi^{2}, \]
\[ \langle \psi^{\dag} \psi^{\dag}\psi \rangle \quad (\delta=0) \quad
{\rm with\ the\ counterterm} \quad (\psi^{\dag})^{2} \psi. \]
The superficial divergence in the function
$\langle \psi^{\dag} \psi vv \rangle$ with $\delta=0$ and the
counterterm $\psi^{\dag} \psi v^{2}$, allowed by the dimension
for the both models, are in fact forbidden by the Galilean
symmetry.\footnote{The arguments based on the Galilean symmetry are
usually applicable to the velocity field governed by the Navier--Stokes
equation, and generally become invalid for synthetic Gaussian velocity
ensembles. It turns out, however, that for a Gaussian ensemble of the type
(\ref{white}) with {\it vanishing} correlation time the Galilean symmetry
of the counterterms indeed takes place. This issue, along with the
consequences of the Galilean invariance for the renormalization, is
discussed in \cite{Alexa} in detail. The proof given there is also
applicable to the models (\ref{ActionA}), (\ref{ActionG}).}
All the remaining terms are present in the corresponding action functionals
(\ref{ActionA}) and (\ref{ActionG}), so that our models are multiplicatively
renormalizable. It is important to stress that, since neither $\psi^{\dag}$
nor $\psi$ are conserved in the full model, the both counterterms
$\psi^{\dag} (v\partial) \psi$  and $\psi^{\dag} (\partial v) \psi$
will be generated by the renormalization procedure. Thus to ensure the
multiplicative renormalizability, one has to include the both such
terms in the original action from the very beginning; hence the
necessity to introduce the two terms with the velocity in (\ref{nabla})
and (\ref{Vertex}).

The Galilean symmetry also requires that the counterterms
$\psi^{\dag}\partial_{t}\psi$ and $\psi^{\dag} (v\partial) \psi $
enter the renormalized action only in the form of the Lagrangian
derivative $\psi^{\dag}\nabla_{t}\psi$, imposing no restriction on
the Galilean invariant term $\psi^{\dag} (\partial v) \psi $.

We thus conclude that the renormalized actions can be written in the forms
\begin{eqnarray}
\S_{A}^{R}(\Phi) = \S_{C}^{R}(\Phi) +
\lambda Z_{4}  (\psi^{\dag})^{2} -
u\mu^{\varepsilon} \lambda Z_{5}  \psi^{\dag} \psi^{3} /3!
\label{ActionAR}
\end{eqnarray}
for the model {\it A} and
\begin{eqnarray}
\S_{G}^{R}(\Phi) = \S_{C}^{R}(\Phi)
+ \frac{g\mu^{\varepsilon/2}\lambda}{2} \left\{ Z_{4}(\psi^{\dagger})^2\psi
- Z_{5} \psi^{\dagger}\psi^2  \right\}
\label{ActionGR}
\end{eqnarray}
for the Gribov model, where the common part of the
renormalized actions is
\begin{eqnarray}
\S_{C}^{R}(\Phi) =  \psi^{\dag} \left\{
- Z_{1} \nabla_{t} + \lambda\left( Z_{2} \partial^{2}- Z_{3}\tau\right)
- a Z_{6} (\partial_{i}v_{i}) \right\} \psi +  \S(\bfv),
\label{ActionCR}
\end{eqnarray}
with $\S(\bfv)$ from (\ref{Sv}).

Here $\lambda$, $\tau$, $g$, $u$ and $a$ are renormalized analogs of the
bare parameters (with the subscripts ``o'') and $\mu$ is the reference mass
scale (additional arbitrary parameter of the renormalized theory). The
renormalization constants $Z_{i}$ absorb the poles in $\varepsilon$ and $\xi$ and
depend on the dimensionless parameters $u$, $w$, $\alpha$ and $a$.
Expressions (\ref{ActionAR})--(\ref{ActionCR}) can be reproduced by the
multiplicative renormalization of the fields $\psi \to \psi Z_{\psi}$,
$\psi^{\dag} \to \psi^{\dag} Z_{\psi^{\dag}}$
and the parameters:
\begin{eqnarray}
g_{0} = g \mu^{\varepsilon/2} Z_{g}, \quad u_{0} = g \mu^{\varepsilon} Z_{u}, \quad
w_{0} = w \mu^{\xi} Z_{w}, \nonumber \\
\lambda_{0} = \lambda Z_{\lambda}, \quad
\tau_{0} = \tau Z_{\tau},  \quad  a_{0} = a Z_{\tau}.
\label{Multy}
\end{eqnarray}

Since the last term $\S(\bfv)$ given by (\ref{Sv}) is not renormalized,
the amplitude $D_{0}$ from (\ref{Kraich}) is expressed in renormalized
parameters as $D_{0} = w_{0} \lambda_{0}  = w\lambda \mu^{\xi}$, while
the parameters $m$ and $\alpha$ are not renormalized: $m_{0} = m$,
$\alpha_{0} = \alpha$. Owing to the Galilean symmetry, the both terms
in the covariant derivative $\nabla_{t}$ are renormalized with the same
constant $Z_{1}$, so that the velocity field is not renormalized, either.
Hence the relations
\begin{eqnarray}
Z_{w}Z_{\lambda} =1, \quad Z_{m}= Z_{\alpha} = Z_{v} =1.
\label{RenD}
\end{eqnarray}
Comparison of the expressions (\ref{ActionA})--(\ref{ActionC}) and
(\ref{ActionAR})--(\ref{ActionCR}) gives the following relations between
the renormalization constants $Z_{1}$--$Z_{6}$ and (\ref{Multy}):
\begin{eqnarray}
Z_{1} = Z_{\psi} Z_{\psi^{\dagger}}, \quad Z_{2} = Z_{1}Z_{\lambda}, \quad
Z_{3} = Z_{2} Z_{\tau}, \quad Z_{6} = Z_{1} Z_{a}
\label{ZZ}
\end{eqnarray}
for the both models,
\begin{equation}
Z_{4} = Z_{\lambda} Z_{\psi^{\dagger}}^{2}, \quad
Z_{5} = Z_{\lambda}Z_{u} Z_{\psi}^{3} Z_{\psi^{\dagger}}
\label{ZA}
\end{equation}
for the model {\it A} and
\begin{equation}
Z_{4} = Z_{g} Z_{\lambda} Z_{\psi^{\dagger}}^{2} Z_{\psi},
\quad Z_{5} = Z_{g} Z_{\lambda} Z_{\psi^{\dagger}} Z_{\psi}^{2}
\label{ZG}
\end{equation}
for the Gribov model.
Resolving these relations with respect to the renormalization constants
of the fields and parameters gives
\begin{equation}
Z_{\lambda} = Z_{1}^{-1} Z_{2}, \quad Z_{\tau} = Z_{2}^{-1} Z_{3}, \quad
Z_{a} = Z_{1}^{-1} Z_{6}, \quad
Z_{u} = Z_{1}^{-1} Z_{2}^{-2} Z_{4} Z_{5}
\label{ResoC}
\end{equation}
for the both models (for the Gribov model, where we passed to the coupling
constant $u=g^{2}$ with $Z_{u}=Z_{g}^{2}$),
\begin{eqnarray}
Z_{\psi}= Z_{1}^{1/2}Z_{2}^{1/2}Z_{4}^{-1/2}, \quad
Z_{\psi}^{\dag}= Z_{1}^{1/2}Z_{2}^{-1/2}Z_{4}^{1/2}
\label{ResoA}
\end{eqnarray}
for the model {\it A} and
\begin{eqnarray}
Z_{\psi}= Z_{1}^{1/2}Z_{4}^{-1/2}Z_{5}^{1/2}, \quad
Z_{\psi}^{\dag}= Z_{1}^{1/2}Z_{4}^{1/2}Z_{5}^{-1/2}
\label{ResoG}
\end{eqnarray}
for the Gribov model.

\begin{figure}
\begin{center}
\includegraphics[width=15cm]{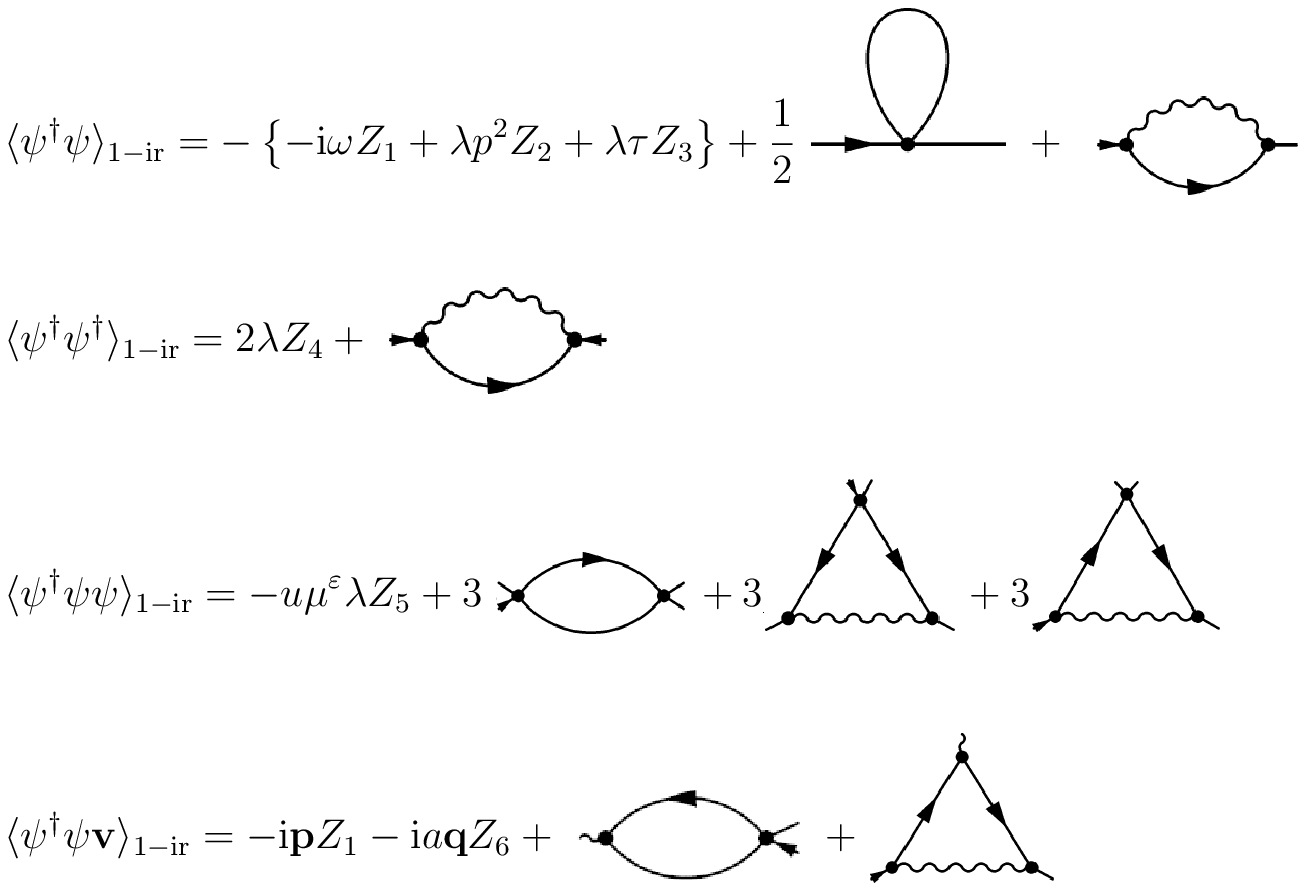}
\caption{\label{fig:DA}
One-loop approximation for the relevant 1-irreducible Green functions
in the model (\protect\ref{ActionA}).}
\end{center}
\end{figure}

\begin{figure}
\begin{center}
\includegraphics[width=15cm]{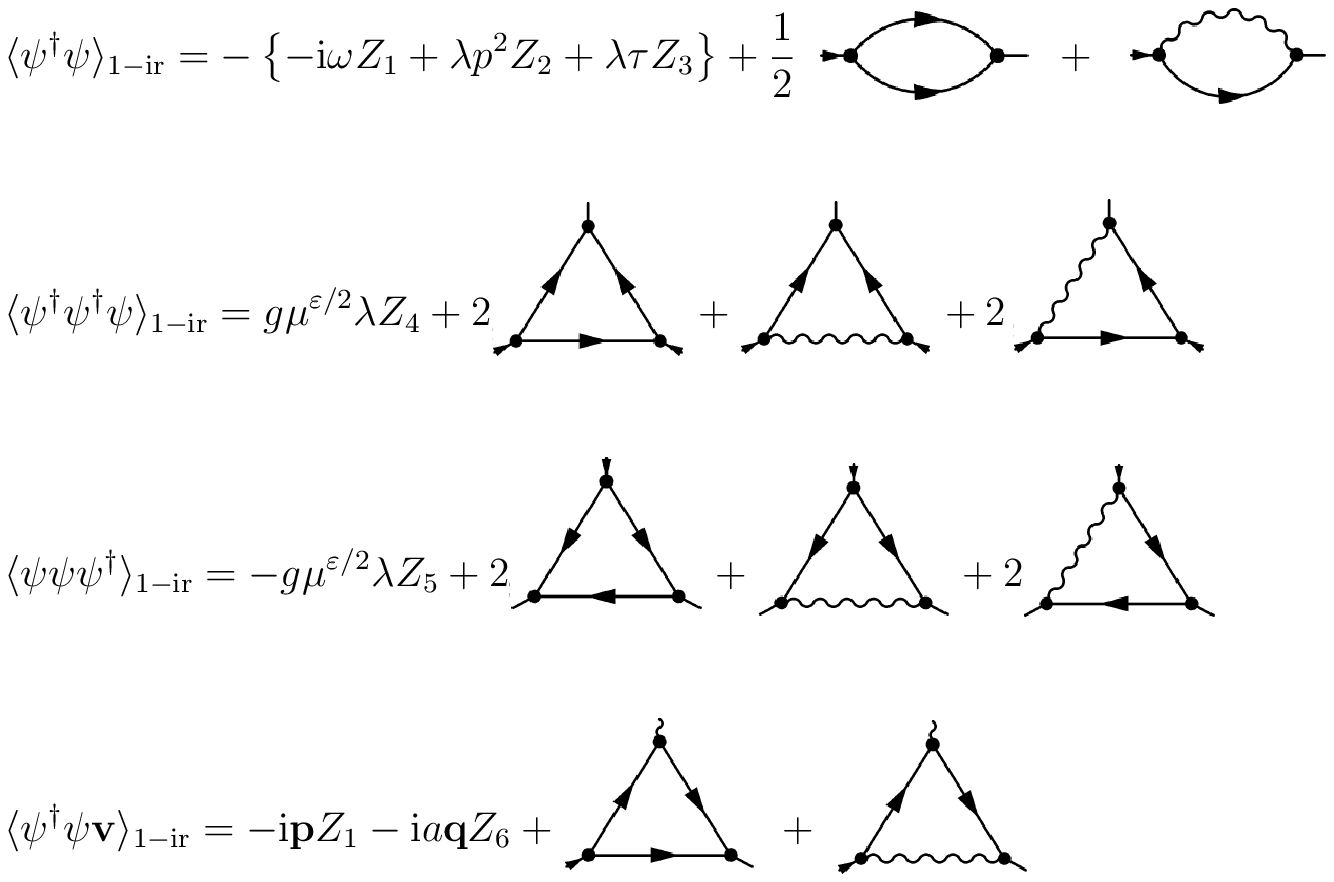}
\caption{\label{fig:DG}
One-loop approximation for the relevant 1-irreducible Green functions
in the model (\protect\ref{ActionG}).}
\end{center}
\end{figure}

The renormalization constants can be found from the requirement that the
Green functions of the renormalized models
(\ref{ActionAR})--(\ref{ActionCR}), when expressed in renormalized
variables, be UV finite (in our case, be finite at $\varepsilon\to0$,
$\xi\to0$). The constants $Z_{1}$--$Z_{6}$ are calculated directly from
the diagrams, then the constants in (\ref{Multy}) are found from
(\ref{ResoC})--(\ref{ResoG}). In order to find the full set of constants,
it is sufficient to consider the 1-irreducible Green functions which involve
superficial divergences; all these functions are listed above.
In the one-loop approximation, they are given in figure~\ref{fig:DA} for the
model {\it A} and in figure~\ref{fig:DG} for the Gribov model.

The solid lines with arrows denote the propagator (\ref{lines}),
the arrow points to the field $\psi^{\dag}$. The solid lines without arrows
correspond to the propagator (\ref{lines2}) and the wavy lines denote the
velocity propagator $\langle vv \rangle_{0}$ specified in (\ref{white}).
The external ends with incoming arrows correspond to the fields
$\psi^{\dag}$, the ends without arrows correspond to $\psi$. The triple
vertex with one wavy line corresponds to the vertex factor (\ref{VertexF}).

The minus signs in front of the bare terms for the functions
$\langle \psi^{\dag} \psi \rangle_{1-ir}$ come from the Dyson equations.
The first and the last diagrams for the functions
$\langle \psi^{\dag} \psi \psi  \rangle_{1-ir}$ and
$\langle \psi^{\dag} \psi^{\dag} \psi \rangle_{1-ir}$ for the Gribov model
have in fact two different forms, related by the mirror reflection,
but they give equal contributions to the renormalization constants and
are accounted by the factors of 2.
What is more, the last diagrams for those functions involve closed circuits
of retarded propagators $\langle \psi \psi^{\dag} \rangle_{0}$ and therefore
vanish identically. For the same reason, the last diagrams in the functions
$\langle \psi^{\dag} \psi \psi  \rangle_{1-ir}$ and for the model {\it A}
and $\langle \psi^{\dag}\psi {\bf v} \rangle_{1-ir}$ for the both models
also vanish.

All the diagrammatic elements should be expressed in renormalized variables
using the relations (\ref{Multy})--(\ref{ZZ}). In the one-loop approximation,
the $Z$'s in the bare terms should be taken in the first order in $u= g^{2}$
and $w$, while in the diagrams they should simply be replaced with unities,
$Z_{i} \to 1$. Thus the passage to renormalized variables in the diagrams
is achieved by the simple substitutions
$\lambda_{0} \to \lambda$, $\tau_{0} \to \tau$,
$g_{0} \to g\mu^{\varepsilon/2}$ and $w_{0} \to w\mu^{\xi}$.

In practical calculations, we used the minimal subtraction (MS) scheme,
in which the renormalization constants have the forms $Z_{i}=1+\,$ only
singularities in $\varepsilon$ and $\xi$, with the coefficients depending
on the completely dimensionless renormalized parameters $u$, $w$, $a$ and
$\alpha$. The one-loop calculation is rather similar to that for the
incompressible case ($\alpha=0$), discussed in \cite{AIK} in detail
(see also \cite{AHH,Alexa}), and here we only give the results:
\begin{eqnarray}
Z_{1} = 1, \quad  Z_{2} = 1 - \frac{w}{4\xi}(3+\alpha), \quad
Z_{3} = 1 + \frac{u}{\varepsilon}, \nonumber \\
Z_{4} = 1 - \frac{w}{\xi} \alpha (a-1)^{2} , \quad
Z_{5} = 1 + \frac{3u}{\varepsilon} - \frac{3w}{\xi} \alpha a^{2},
\nonumber \\
Z_{6} = 1 + \frac{u}{\varepsilon}\, \frac{(4a-1)}{4a}
\label{ZoA}
\end{eqnarray}
for the model {\it A} and
\begin{eqnarray}
Z_{1} = 1 + \frac{u}{4\varepsilon}, \quad
Z_{2} = 1 + \frac{u}{8\varepsilon} - \frac{w}{4\xi}(3+\alpha), \quad
Z_{3} = 1 + \frac{u}{\varepsilon}, \nonumber \\
Z_{4} = 1 + \frac{u}{\varepsilon} - \frac{w}{\xi} \alpha (a-1)^{2} , \quad
Z_{5} = 1 + \frac{u}{\varepsilon} - \frac{w}{\xi} \alpha a^{2}, \nonumber \\
Z_{6} = 1 + \frac{u}{\varepsilon}\, \frac{(4a-1)}{8a}
\label{ZoG}
\end{eqnarray}
for the Gribov model, with higher-order corrections in the couplings.
To simplify the resulting expressions, we passed in (\ref{ZoA})
and (\ref{ZoG}) to the new parameters
\[ u \to u/16\pi^2, \quad w \to w/16\pi^2; \]
here and below they are denoted by the same symbols $u$ and $w$.

For $w=0$, the well-known one-loop results for the models (\ref{actionA}),
(\ref{actionG}) are reproduced (up to the notation). For the incompressible
case ($\alpha=0$) expressions (\ref{ZoG}) coincide with the results obtained
earlier in \cite{AIK}. Furthermore, the first-order expressions (\ref{ZoG})
satisfy exact relations
\begin{eqnarray}
Z_{i} (a) = Z_{i} (1-a) \quad {\rm for} \quad i=1,2,3; \nonumber \\
Z_{4} (a) = Z_{5} (1-a), \quad
Z_{1} (a) -a Z_{6}(a) = (1-a) Z_{6}(1-a),
\label{GSymm}
\end{eqnarray}
which follow from the symmetries (\ref{symm}), (\ref{symmG}) of the Gribov
model. These relations will play an important role in the analysis of the
fixed points of the model (\ref{ActionG}) in the next sections.

\section{RG functions and RG equations} \label{sec:RGE}

Let us recall an elementary derivation of the RG equations; detailed
exposition can be found in monographs \cite{Zinn,Book3}.
The RG equations are written for the renormalized Green functions
$G^{R} =\langle \Phi\cdots\Phi\rangle_{R}$, which differ from the original
(unrenormalized) ones $G =\langle \Phi\cdots\Phi\rangle$ only by
normalization (due to rescaling of the fields) and choice of parameters,
and therefore can equally be used for analyzing the critical behaviour.
The relation $\S^{R} (\Phi,e,\mu) = \S(\Phi,e_{0})$ between the bare
(\ref{ActionA})--(\ref{ActionC}) and renormalized
(\ref{ActionAR})--(\ref{ActionCR}) action functionals results
in the relations
\begin{equation}
G(e_{0},\dots) = Z_{\psi}^{N_{\psi}} Z_{\psi^{\dagger}}^{N_{\psi^{\dagger}}}
G^{R}(e,\mu,\dots).
\label{multi}
\end{equation}
between the Green functions. Here, as usual, $N_{\psi}$ and
$N_{\psi^{\dagger}}$ are the numbers of corresponding fields
entering into $G$ (we recall that in our models $Z_{v}=1$);
$e_{0}=\{\lambda_{0}, \tau_{0}, u_{0}, w_{0}, a_{0}, m_{0}, \alpha_{0} \}$
is the full set of bare parameters and
$e=\{ \lambda, \tau, u, w, a, m, \alpha  \}$ are their renormalized
counterparts (we recall that $\alpha_{0}=\alpha$ and $m_{0}=m$);
the dots stand for the other arguments
(times/frequencies and coordinates/momenta).

We use $\widetilde{\cal D}_{\mu}$ to denote the differential operation
$\mu\partial_{\mu}$ for fixed $e_{0}$ and operate on both sides of the
equation (\ref{multi}) with it. This gives the basic RG differential
equation:
\begin{equation}
\left\{ {\cal D}_{RG} + N_{\psi} \gamma_{\psi} +
N_{\psi^{\dag}} \gamma_{\psi^{\dag}} \right\}
\,G^{R}(e,\mu,\dots) = 0,
\label{RG1}
\end{equation}
where ${\cal D}_{RG}$ is the operation $\widetilde{\cal D}_{\mu}$
expressed in the renormalized variables:
\begin{equation}
{\cal D}_{RG}\equiv {\cal D}_{\mu} + \beta_{u}\partial_{u} +
\beta_{w}\partial_{w}  + \beta_{a}\partial_{a} -
\gamma_{\lambda}{\cal D}_{\lambda} - \gamma_{\tau}{\cal D}_{\tau}.
\label{RG2}
\end{equation}
Here we have written ${\cal D}_{x}\equiv x\partial_{x}$ for any variable
$x$, the anomalous dimensions $\gamma$ are defined as
\begin{equation}
\gamma_{F}\equiv \Dm \ln Z_{F} \quad {\rm for\ any\ quantity} \ F,
\label{RGF1}
\end{equation}
and the $\beta$ functions for the dimensionless couplings $u$, $w$ and
$a$ are
\begin{eqnarray}
\beta_{u} \equiv \widetilde {\cal D}_{\mu} u = u\, (-\varepsilon-\gamma_{u}),
\nonumber \\
\beta_{w} \equiv \widetilde {\cal D}_{\mu} w = w\,(-\xi-\gamma_{w}),
\nonumber \\
\beta_{a} \equiv \widetilde {\cal D}_{\mu} a = -a\gamma_{a},
\label{betagw}
\end{eqnarray}
where the second equalities come from the definitions and the
relations (\ref{Multy}). The fourth $\beta$ function
\begin{eqnarray}
\beta_{\alpha}=\widetilde{\cal D}_{\mu}\alpha=-\alpha\gamma_{\alpha}
\label{Bal}
\end{eqnarray}
vanishes identically due to (\ref{exi}) and for this reason does not
appear in the subsequent relations.

The anomalous dimension corresponding to a given renormalization constant
$Z_{F}$ is readily found from the relation
\begin{equation}
\gamma_{F} = \left( \beta_{u}\partial_{u}+\beta_{w}\partial_{w}
+\beta_{a}\partial_{a} \right)
\ln Z_{F} \simeq  - \left(\varepsilon\D_{u}+\xi\D_{w}\right) \ln Z_{F}.
\label{GfZ}
\end{equation}
In the first relation, we used the definition (\ref{RGF1}), expression
(\ref{RG2}) for the operation $\Dm$ in renormalized variables, and the
fact that the $Z$'s depend only on the completely dimensionless coupling
constants $u$, $w$ and $a$. In the second (approximate) relation, we only
retained the leading-order terms in the $\beta$ functions (\ref{betagw}),
which is sufficient for the first-order approximation. The leading-order
expressions (\ref{ZoA}), (\ref{ZoG}) for the renormalization constants have
the form
\begin{equation}
Z_{F} = 1 + \frac{u}{\varepsilon} A_{F}(a,\alpha) + \frac{w}{\xi} B_{F}(a,\alpha).
\label{Zf}
\end{equation}
Substituting (\ref{Zf}) into (\ref{GfZ}) leads to the final UV finite
expressions for the anomalous dimensions:
\begin{equation}
\gamma_{F} = - u A_{F}(a,\alpha) - w B_{F}(a,\alpha)
\label{gift}
\end{equation}
for any constant $Z_{F}$. This gives
\begin{eqnarray}
\gamma_{1} = 0, \quad \gamma_{2} = w (3+\alpha)/4, \quad
\gamma_{3} = -u , \quad \gamma_{4} =w \alpha (a-1)^{2} , \nonumber \\
\gamma_{5} =  -3u +3 w \alpha a^{2}, \quad
\gamma_{6} =u(1-4a)/4a
\label{anomA}
\end{eqnarray}
for the model {\it A} and
\begin{eqnarray}
\gamma_{1} = -u/4 , \quad \gamma_{2} = -u/8 + w (3+\alpha)/4, \quad
\gamma_{3} = -u/2 , \nonumber \\
\gamma_{4} = -u + w \alpha (a-1)^{2}, \quad
\gamma_{5} = -u + w \alpha a^{2}, \quad
\gamma_{6} =u(1-4a)/8a
\label{anomG}
\end{eqnarray}
for the Gribov model.

The multiplicative relations (\ref{ResoC})--(\ref{ResoG})
between the renormalization constants result in the linear relations
between the corresponding anomalous dimensions:
\begin{eqnarray}
\gamma_{\lambda} =  \gamma_{2} -\gamma_{1}, \quad
\gamma_{\tau} =  \gamma_{3} -\gamma_{2} , \nonumber \\
\gamma_{a} =  \gamma_{6} -\gamma_{1}, \quad
\gamma_{u} = -\gamma_{1}- 2\gamma_{2} +\gamma_{4} + \gamma_{5}
\label{aesoC}
\end{eqnarray}
for the both models,
\begin{eqnarray}
2\gamma_{\psi}= \gamma_{1} +\gamma_{2} -\gamma_{4}, \quad
2\gamma_{\psi}^{\dag}= \gamma_{1} - \gamma_{2} + \gamma_{4}
\label{aesoA}
\end{eqnarray}
for the model {\it A} and
\begin{eqnarray}
2\gamma_{\psi}= \gamma_{1}- \gamma_{4} + \gamma_{5}, \quad
2\gamma_{\psi}^{\dag}= \gamma_{1}+ \gamma_{4} -  \gamma_{5}
\label{aesoG}
\end{eqnarray}
for the Gribov model. Along with (\ref{anomA}), (\ref{anomG}), these
relations give the final
first-order explicit expressions for the anomalous dimensions of the
fields and parameters. The exact relations (\ref{RenD}) result in
\begin{eqnarray}
\gamma_{w} =-\gamma_{\lambda}, \quad
\gamma_{m} =\gamma_{\alpha} =\gamma_{v} = 0,
\label{exi}
\end{eqnarray}
while the relations (\ref{GSymm}) give
\begin{eqnarray}
\gamma_{i} (a) = \gamma_{i} (1-a) \quad {\rm for} \quad i=1,2,3;
\nonumber \\
\gamma_{4} (a) = \gamma_{5} (1-a), \quad
\gamma_{1} (a) -a \gamma_{6}(a) = (1-a) \gamma_{6}(1-a)
\label{SymmG}
\end{eqnarray}
for the Gribov model.

\section{Attractors of the RG equations and scaling regimes} \label{sec:FPS}

It is well known that possible asymptotic regimes of a renormalizable field
theoretic model is determined by the asymptotic behaviour of the system of
ordinary differential equations for the so-called invariant (running)
coupling constants
\begin{eqnarray}
\D_s \bar g_{i}(s,g) = \beta_{i} (\bar g), \quad \bar g_{i}(1,g) = g_{i},
\label{Odri}
\end{eqnarray}
where $s=k/\mu$, $k$ is the momentum,
$g= \{g_{i}\}$ is the full set of coupling constants and
$\bar g_{i}(s,g)$ are the corresponding invariant variables. As a rule,
the IR ($s\to0$) and UV ($s\to\infty$) behaviour of such system is
determined by fixed points $g_{i*}$. The coordinates of possible fixed
points are found from the requirement that all the $\beta$ functions vanish:
\begin{eqnarray}
\beta_{i} (g_{*}) =0,
\label{fp}
\end{eqnarray}
while the type of a given fixed point is determined by the matrix
\begin{equation}
\Omega_{ij} = \partial\beta_{i}/\partial g_{j} |_{g=g^*}:
\label{OmegaDef}
\end{equation}
for an IR attractive fixed points (which we are interested in here) the
matrix $\Omega$ is positive, that is, the real parts of all its eigenvalues
are positive. In our models, the fixed points for the full set of couplings
$u$, $w$, $a$, $\alpha$ should be determined by the equations
\begin{equation}
\beta_{u,w,a,\alpha} (u_{*},w_{*},a_{*},\alpha_{*}) = 0,
\label{points}
\end{equation}
with the $\beta$ functions defined in the preceding section. However, in our
models the attractors of the system (\ref{Odri}) involve, in general,
two-dimensional surfaces in the full four-dimensional space of couplings.
First, the function (\ref{Bal}) vanishes identically, so that the
equation $\beta_{\alpha}=0$ gives no restriction on the parameter $\alpha$.
It is then convenient to consider the attractors of the system (\ref{Odri})
in the three-dimensional space $u$, $w$, $a$; their coordinates, matrix
(\ref{OmegaDef}) and the critical exponents will, in general, depend on
the free parameter $\alpha$. What is more, in this reduced space the
attractors will be not only fixed points, but also lines of fixed points,
which can be conveniently parametrized by the coupling $a$. Although the
general pattern of the attractors appears rather similar for the both
models, it is instructive to discuss them separately.

\subsection{Scaling regimes for the Gribov model} \label{sec:GPS}

The one-loop expressions for the $\beta$ functions in the model
(\ref{ActionGR}), (\ref{ActionCR}) are easily derived from the definitions
(\ref{betagw}), relations (\ref{aesoC}), (\ref{aesoG}) and (\ref{exi}),
and explicit expressions (\ref{anomG}):
\begin{eqnarray}
\beta_{u} = u \left[ -\varepsilon+ 3u/2 + w(3+\alpha)/2 -w\alpha f(a) \right],
\nonumber \\
\beta_{w} = w \left[ -\xi +u/8 + w(3+\alpha)/4 \right],
\nonumber \\
\beta_{a} = u(2a-1)/8,
\label{betaG}
\end{eqnarray}
where the function $f(a)=a^{2}+(a-1)^{2}$ achieves the minimum value
$f(1/2) =1/2$ at $a=1/2$. For the set (\ref{betaG}) the equations
(\ref{points}) have the following four solutions:

(1) The line of Gaussian (free) fixed points: $u_{*}=w_{*}=0$, $a_{*}$
arbitrary.

(2) The point $w_{*}=0$, $u_{*}=2\varepsilon/3$, $a_{*}=1/2$, corresponding
to the pure Gribov model (turbulent advection is irrelevant).

(3) The line of fixed points
\begin{equation}
u_{*}=0, \quad w_{*}=4\xi/(3+\alpha), \quad  a_{*} \ {\rm arbitrary},
\label{line3}
\end{equation}
corresponding to the passively advected scalar without self-interaction.

(4) The most nontrivial fixed point, corresponding to the new regime
(universality class), both the advection and the self-interaction are
relevant:
\begin{equation}
u_{*} = \frac{4\,[\varepsilon(3+\alpha)-6\xi]}{3(5+2\alpha)}, \quad
w_{*} = \frac{8\,[-\varepsilon/4+3\xi]}{3(5+2\alpha)}, \quad a_{*}=1/2.
\label{wu4}
\end{equation}
We recall that $\alpha$ is treated as a free parameter, which the
coordinates of the fixed points can depend on.
For $a=1/2$, the relations (\ref{aesoC}) and (\ref{SymmG}) give
$\gamma_{a} = \gamma_{6}-\gamma_{1}=0$, so that $\beta_{a} =-a \gamma_{a}$
vanishes to all orders of the perturbation theory, irrespective of the
values of the parameters $u$ and $w$. It then follows that for the fixed
points (2) and (4), the value $a_{*}=1/2$ is in fact exact and has no
higher-order corrections in $\varepsilon$ and $\xi$.
The expression for $w_{*}$
for the point (3), which corresponds to the exactly soluble Kraichnan model,
is also exact, while the expressions for $u_{*}$ and $w_{*}$ for the
fourth point are the first-order terms of infinite double expansions
in $\varepsilon$ and $\xi$.

\begin{figure}
\begin{center}
\includegraphics[width=11cm]{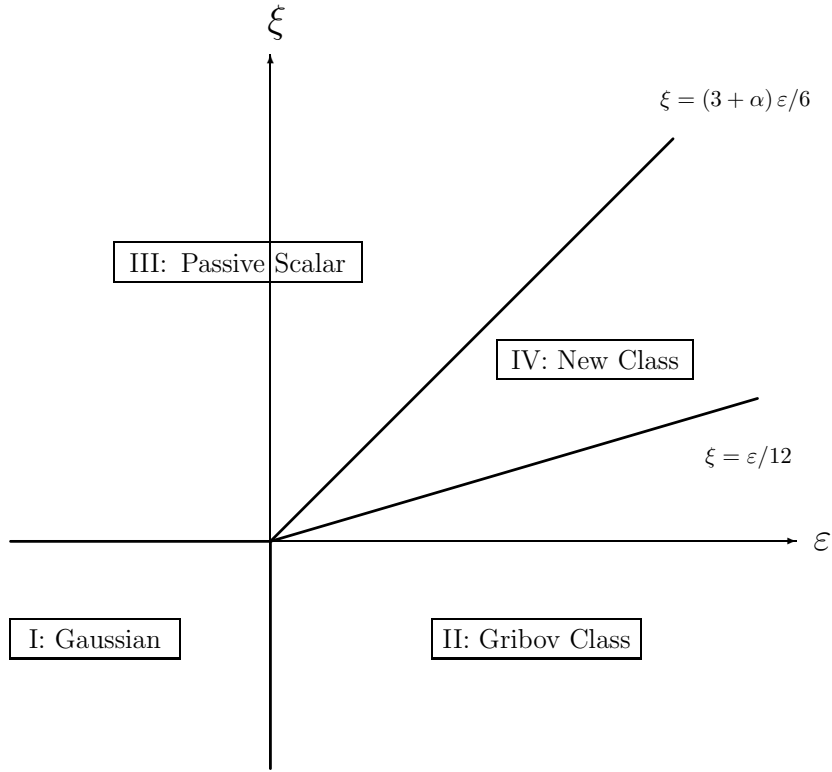}
\caption{\label{fig:patt}
Regions of admissibility of the fixed points in
the model (\protect\ref{ActionG}).}
\end{center}
\end{figure}

Admissible fixed point must be IR attractive and satisfy the conditions
$u_{*}>0$, $w_{*}>0$, which follow from the physical meaning of these
parameters. In the one-loop approximations, these two requirements in fact
coincide for all the fixed points. In figure~\ref{fig:patt} we show
the regions of IR stability of the fixed points (1)--(4) in the
$\varepsilon$--$\xi$ plane. For the points (1)--(3), the matrix
(\ref{OmegaDef}) is either diagonal
or triangular, and the stability is in fact determined by its diagonal
elements $\Omega_{i} = \partial\beta_{i}/\partial g_{i}$.

For the point (1) one has
\[ \Omega_{u} = -\varepsilon, \quad  \Omega_{w} = -\xi, \quad \Omega_{a} = 0, \]
so it is admissible for $\varepsilon<0$, $\xi<0$ (region I in
fig.~\ref{fig:patt}). Vanishing of the element $\Omega_{a}$ reflects the
fact that the parameter $a_{*}$ for the point I is arbitrary, or, in other
words, the point is degenerate.
For the point (2) one finds
\[ \Omega_{u} = \varepsilon, \quad
\Omega_{w} = -\xi+u_{*}/8 = -\xi+\varepsilon/12,
\quad \Omega_{a} = u_{*}/4 = \varepsilon/6, \]
so it is admissible for $\varepsilon>0$, $\xi<\varepsilon/12$
(region II in fig.~\ref{fig:patt}).
For the point (4) the elements $\partial\beta_{a}/\partial u$,
$\partial\beta_{a}/\partial w$ vanish, so that the matrix $\Omega$ appears
block-triangular and the eigenvalue $\Omega_{a} = u_{*}/4$ decouples.
The eigenvalues of the remaining $2\times2$ matrix for the couplings
$u$, $w$ look rather complicated, but the determinant is simple and equals
$w_{*}u_{*} (\alpha+5/2)$. This expression suggests that the matrix is
positive for $u_{*}>0$, $w_{*}>0$, and careful analysis shows that this
is indeed so. From the explicit expressions (\ref{wu4}) we conclude that
the point (4) is admissible for $\xi>\varepsilon/12$,
$\xi< (3+\alpha) \varepsilon/6$ (region IV in fig.~\ref{fig:patt}).

The remaining sector III in fig.~\ref{fig:patt} is specified by the
inequalities
\begin{equation}
\xi > 0, \qquad (3+\alpha) \varepsilon -6\xi<0,
\label{sector3}
\end{equation}
while the eigenvalues of the matrix (\ref{OmegaDef}) at the point (3)
are as follows
\begin{equation}
\Omega_{u} =  -\varepsilon+2\xi- \frac{4\alpha\xi}{(3+\alpha)}\, f(a_{*}),
\quad \Omega_{w} = \xi, \quad \Omega_{a} = 0
\label{omga3}
\end{equation}
with $f(a)$ from (\ref{betaG}). The first inequality in (\ref{sector3})
coincides with $\Omega_{w}>0$, while the second can be rewritten as
\begin{equation}
\Omega_{u}({a_{*}=1/2}) = -\varepsilon+2\xi-
\frac{4\alpha\xi}{(3+\alpha)}\, f(1/2) >0.
\label{omga32}
\end{equation}
Comparison of equations (\ref{omga3}) and (\ref{omga32}) shows that the
condition $\Omega_{u}>0$ can be satisfied only in the region III, provided
the coordinate $a_{*}$ satisfies the inequality
\begin{equation}
(a_{*}-1/2)^{2} \le \frac{1}{8\alpha\xi}\,
\bigl[ 6\xi- (3+\alpha)\varepsilon \bigr]
\label{kut}
\end{equation}
(it is important here that $f(1/2)=1/2$ is the minimum value of $f$).
Thus we conclude that the third regime, for which the self-interaction is
irrelevant, is admissible in the sector III, specified by the inequalities
(\ref{sector3}). The corresponding attractor has the form of an interval
on the line (\ref{line3}), centered at $a_{*}=1/2$ and defined by the
inequality (\ref{kut}). This interval becomes infinite for $\alpha=0$
and shrinks to the single point $a_{*}=1/2$ for $\alpha\to\infty$.
It should be noted, however, that for the purely transverse velocity
field ($\alpha=0$) the term with the coefficient $a_{0}$ in (\ref{nabla})
and all the subsequent expressions vanishes, and this parameter in fact
disappears from the model.

In the one-loop approximation (\ref{betaG}), all the boundaries between the
regions of admissibility are given by straight rays; there are neither gaps
nor overlaps between the different regions. Due to higher-order corrections,
the boundaries between the regions II and IV and between the regions III and
IV can change and become curved. However, it can be argued that no gaps nor
overlaps can appear between them to all orders. The similar situation was
encountered earlier in the crossover between the long-range and short-range
regimes of the Gribov model; see \cite{Levy}. Here we only note that,
in the present case, it is important that the models with $u=0$ or $w=0$ are
``closed with respect to renormalization'' so that the functions $\beta_{u}$
for $w=0$ and $\beta_{w}$ for $u=0$ coincide with the $\beta$ functions of
the Gribov and rapid-change models to all orders of the perturbation theory.

It remains to note that the location of the boundary between the regions
III and IV depends on the parameter $\alpha$. For the $\alpha=0$, it is
given by the ray $\xi=\varepsilon/2$, in agreement with the result, derived
earlier in \cite{AIK} for the incompressible case. When $\alpha$ increases,
the ray rotates counter clockwise, and in the limit $\alpha\to\infty$ it
approaches the vertical ray $\varepsilon=0$, $\xi>0$.

\subsection{Scaling regimes for the model {\it A}} \label{sec:GPA}

The one-loop expressions for the $\beta$ functions in the model
(\ref{ActionAR}), (\ref{ActionCR}) are easily derived from the definitions
(\ref{betagw}), relations (\ref{aesoC}), (\ref{aesoA}) and (\ref{exi}),
and explicit expressions (\ref{anomA}):
\begin{eqnarray}
\beta_{u} = u \left[ -\varepsilon+ 3u + w(3+\alpha)/2 -w\alpha f(a) \right],
\nonumber \\
\beta_{w} = w \left[ -\xi + w(3+\alpha)/4 \right],
\nonumber \\
\beta_{a} = u(4a-1)/4,
\label{betaA}
\end{eqnarray}
where the function $f(a)=3a^{2}+(a-1)^{2}$ achieves the minimum value
$f(1/4) =3/4$ at $a=1/4$.

(1) The line of Gaussian (free) fixed points: $u_{*}=w_{*}=0$, $a_{*}$
arbitrary. IR attractive for $\varepsilon<0$, $\xi<0$.

(2) The point $w_{*}=0$, $u_{*}=\varepsilon/3$, $a_{*}=1/4$, corresponding
to the pure model {\it A} (turbulent advection is irrelevant). This point
is IR attractive for $\xi<0$, $\varepsilon<0$.

(3) The line of fixed points
\begin{equation}
u_{*}=0, \quad w_{*}=4\xi/(3+\alpha), \quad  a_{*} \ {\rm arbitrary},
\label{line4}
\end{equation}
corresponding to the passively advected scalar without self-interaction.
It involves the interval
\begin{equation}
(a_{*}-1/4)^{2} < \frac{1}{16\alpha\xi} \left[ -\varepsilon(3+\alpha) +
\xi(6-\alpha)  \right],
\label{line5}
\end{equation}
which is IR attractive for $\xi>0$, $\varepsilon< \xi (6-\alpha)/(3+\alpha)$.
This interval becomes infinite for $\alpha\to0$ and tends to the finite
value $-(\varepsilon+\xi)/16\xi$ for $\alpha\to\infty$ (note that the right hand
side of the inequality (\ref{line5}) is positive within the region of
IR stability).

(4) The fixed point, corresponding to the new universality class, where
both the advection and the self-interaction are relevant:
\begin{equation}
w_{*}= 4\xi/(3+\alpha), \quad u_{*} = \frac{ \varepsilon (3+\alpha) -
\xi(6-\alpha)} {3(3+\alpha)}, \quad a_{*}=1/4.
\label{wu44}
\end{equation}
This point is IR attractive for $\xi>0$,
$\varepsilon> \xi (6-\alpha)/(3+\alpha)$.

\begin{figure}
\begin{center}
\includegraphics[width=11cm]{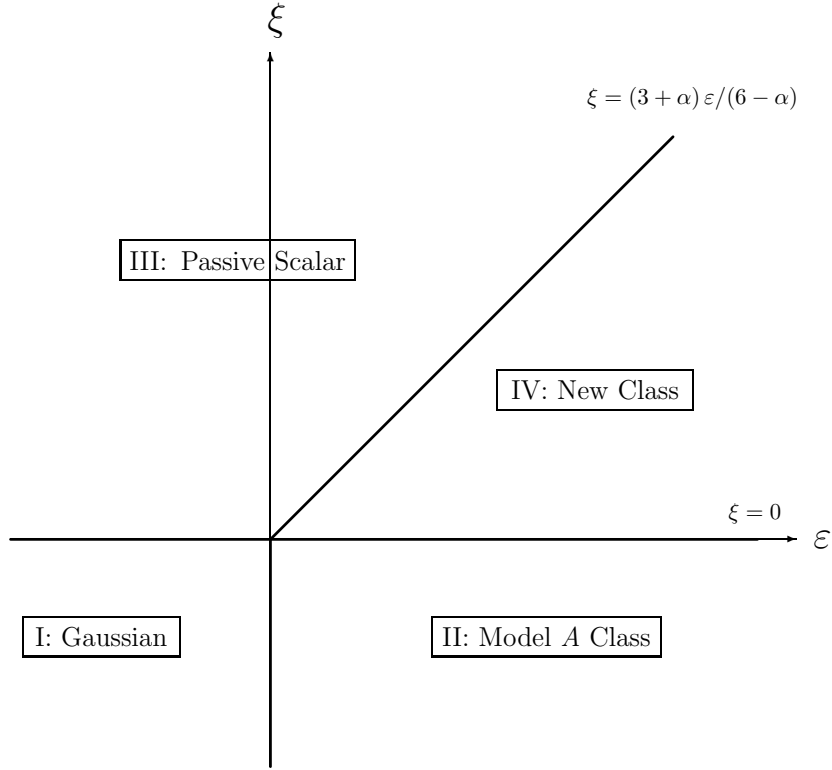}
\caption{\label{fig:pattA} Regions of admissibility of the fixed points
in the model (\protect\ref{ActionA}). The case $\alpha<6$ is shown;
for $\alpha>6$ the boundary between III and IV lies in the left
upper quadrant.}
\end{center}
\end{figure}

When $\alpha$ increases, the boundary between the regions of stability of
the regimes (3) and (4) rotates in the upper half-plane $\varepsilon$--$\xi$
counter clockwise from the ray $\xi=\varepsilon/2$ ($\alpha\to0$) to
$\xi=-\varepsilon$ ($\alpha\to\infty$).
The regions of IR stability of the fixed points (1)--(4) in the
$\varepsilon$--$\xi$ plane are shown in figure~\ref{fig:pattA}
for some value of $\alpha<6$, when the boundary between the regions
III and IV lies in the right upper quadrant; for $\alpha>6$ it moves
to the left upper quadrant.

\section{Critical scaling and critical dimensions} \label{sec:DimeNS}

Existence of IR attractors of the RG equations implies existence of
self-similar (scaling) behaviour of the Green functions in the IR range.
In this critical scaling all the IR irrelevant parameters ($\lambda$,
$\mu$ and the coupling constants) are fixed and the IR relevant parameters
(times/frequencies, coordinates/momenta, $\tau$ and the fields) are dilated.
In dynamical models, critical dimensions $\Delta_{F}$ of the IR relevant
quantities $F$ are given by the relations
\begin{eqnarray}
\Delta_{F} = d^{k}_{F}+ \Delta_{\omega} d^{\omega}_{F} + \gamma_{F}^{*},
\qquad  \Delta_{\omega}=2 -\gamma_{\lambda}^{*},
\label{dim}
\end{eqnarray}
with the normalization condition $\Delta_{k} = 1$; see e.g. \cite{Book3}
for more detail. Here $d^{k,\omega}_{F}$ are the canonical dimensions of
$F$, given in table~\ref{table1}, and $\gamma_{F}^{*}$ is the value of the
corresponding anomalous dimension (\ref{RGF1}) at the fixed point:
$\gamma_{F}^{*} = \gamma_{F} (u_{*},w_{*},a_{*})$. This gives
$\Delta_{\psi} = d/2+ \gamma_{\psi}^{*}$,
$\Delta_{\psi^{\dag}} = d/2+ \gamma_{\psi^{\dag}}^{*}$ for the Gribov model,
$\Delta_{\psi} = d/2-1+ \gamma_{\psi}^{*}$,
$\Delta_{\psi^{\dag}} = d/2+1+ \gamma_{\psi^{\dag}}^{*}$ for the model
{\it A} and $\Delta_{\omega} = 2 + \gamma_{\tau}^{*}$ for the both models.
Substituting the coordinates of the fixed points from sections~\ref{sec:GPS}
and~\ref{sec:GPA} into the explicit one-loop expressions (\ref{anomA}),
(\ref{anomG}) for the anomalous dimensions and using the exact relations
(\ref{aesoC})--(\ref{aesoG}), one obtains the leading-order expressions
for the critical dimensions. They are summarized in tables~\ref{tableG}
and~\ref{tableA} for the Gribov model and model {\it A}, respectively.

\begin{table}
\caption{Critical dimensions of the fields and parameters in the
model (\protect\ref{ActionG}), (\protect\ref{ActionC}).}
\label{tableG}
\begin{tabular}{ccccc}
\br
{} & FP1 & FP2 & FP3 & FP4 \\ \br
$\Delta_{\omega}$ & 2  & $2-\frac{\varepsilon}{12}$  & $2-\xi$ & $2-\xi$
\\ \mr
$\Delta_{\psi}$ & $d/2$  & $2-\frac{7\varepsilon}{12}$  &
$2-\frac{\varepsilon}{2}+\frac{2\xi\alpha(2a-1)}{(3+\alpha)}$ &
$2- \frac{\varepsilon(18+7\alpha)-6\xi}{6(5+2\alpha)}$ \\ \mr
$\Delta_{\psi^{\dag}}$ & $d/2$  & $2-\frac{7\varepsilon}{12}$  &
$2-\frac{\varepsilon}{2}-\frac{2\xi\alpha(2a-1)}{(3+\alpha)}$    &
$2- \frac{\varepsilon(18+7\alpha)-6\xi}{6(5+2\alpha)}$ \\ \mr
$\Delta_{\tau}$ & 2  & $2-\frac{\varepsilon}{4}$ & $2-\xi$  &
$2- \frac{\varepsilon(3+\alpha) +3\xi(3+2\alpha)} {3(5+2\alpha)}$ \\ \br
\end{tabular}
\end{table}

\begin{table}
\caption{Critical dimensions of the fields and parameters in the
model (\protect\ref{ActionA}), (\protect\ref{ActionC}).}
\label{tableA}
\begin{tabular}{ccccc}
\br
{} & FP1 & FP2 & FP3 & FP4 \\ \br
$\Delta_{\omega}$ & 2  & 2  & $2-\xi$ & $2-\xi$ \\ \mr
$\Delta_{\psi}$ & $d/2-1$  & $1-\frac{\varepsilon}{2}$ &
$1+ \frac{\xi-\varepsilon}{2} - \frac{2\xi\alpha(a-1)^{2}} {(3+\alpha)}$ &
$1 - \frac{4\varepsilon(3+\alpha)+\xi(5\alpha-12)}{3(3+\alpha)}$ \\ \mr
$\Delta_{\psi^{\dag}}$ & $d/2+1$  & $3-\frac{\varepsilon}{2}$ &
$3- \frac{\xi+\varepsilon}{2} + \frac{2\xi\alpha(a-1)^{2}} {(3+\alpha)}$ &
$3 - \frac{4\varepsilon(3+\alpha)-\xi(5\alpha-12)}{3(3+\alpha)}$ \\ \mr
$\Delta_{\tau}$ & 2  & $2-\frac{\varepsilon}{3}$  &  $2-\xi$ &
$2- \frac{\varepsilon(3+\alpha)+\xi(3+4\alpha)}{(3+\alpha)}$ \\ \br
\end{tabular}
\end{table}

The results for the Gaussian points (1) are exact; all the results for the
fixed points (2) have corrections of order $\varepsilon^{2}$ and higher.
The result
$\Delta_{\omega}=2-\xi$ for the fixed points (3) and (4) is also exact,
as follows from the general relations $\gamma_{\lambda}=\gamma_{w}$ in
(\ref{exi}) and $\Delta_{\omega}=2-\gamma_{\lambda}^{*}$ in (\ref{dim}),
and the identity $\gamma_{w}^{*}= \xi$, which is a consequence of the
fixed-point equation $\beta_{w}=0$ with $\beta_{w}$ from (\ref{betagw})
for any fixed point with $w_{*}\ne 0$. The other results for the fixed
points (4) have higher-order corrections in $\varepsilon$ and $\xi$.

The result $\Delta_{\tau}=2-\xi$ for the fixed points (3) is also exact:
for $u=0$, the 1-irreducible function $\langle \psi^{\dag} \psi \rangle$
is given by the only one-loop diagram, which gives rise only to the
counterterm $\psi^{\dag}\partial^{2} \psi$; cf. \cite{JphysA} for the pure
Kraichnan's model. Hence the exact relations $Z_{1}=Z_{3}=1$ for $u=0$.
Then from the relations (\ref{ZZ}) and (\ref{exi})
it follows $\gamma_{\tau}=-\gamma_{\lambda}=\gamma_{w}$ which gives
$\gamma_{\tau}^{*}= -\xi$ for the fixed points with $w_{*}\ne 0$.

Fixed point (3) in table~\ref{tableG} illustrates the general fact that
the dimensions $\Delta_{\psi}$ and $\Delta_{\psi^{\dag}}$ for the Gribov
model interchange under the transformation $a\to1-a$, as a consequence of
the exact symmetry (\ref{symm}), (\ref{symmG}), which results in the
relations (\ref{GSymm}) for the renormalization constants.
It remains to note that for any fixed point the critical dimension of the
velocity field is given by the exact relation
$2\Delta_{v}=-\xi+\Delta_{\omega}$ following from the form of its
pair correlation function (\ref{white}), (\ref{Kraich}).

\section{Discussion and conclusions} \label{sec:Conc}

We studied effects of turbulent mixing on the critical behaviour, with
special attention paid to {\it compressibility} of the fluid. Two
representative models of dynamical critical behaviour were considered:
the model {\it A}, which describes relaxational dynamics of a non-conserved
order parameter in an equilibrium critical system, and the strongly
non-equilibrium Gribov model, which describes spreading processes in
a reaction-diffusion system. The turbulent mixing was modelled by the
Kazantsev--Kraichnan ``rapid-change'' ensemble: time-decorrelated Gaussian
velocity field with the power-like spectrum $\propto k^{-d-\xi}$.
The both stochastic problems can be reformulated as multiplicatively
renormalizable field theoretic models, which allows one to apply the field
theoretic renormalization group to the analysis of their IR behaviour.

We showed that, depending on the relation between the spatial dimension
$d$ and the exponent $\xi$, the both models exhibit four different critical
regimes, associated with four possible fixed points of the RG equations.
Three fixed points correspond to known regimes: (1) Gaussian fixed point;
(2) critical behaviour typical of the original model without mixing (that is,
model {\it A} or Gribov model); (3) scalar field without self-interaction,
passively advected by the flow (the nonlinearity in the order parameter in
the original dynamical equations appears unimportant). The most interesting
fourth point corresponds to a new type of critical behaviour (4), in which
the nonlinearity and turbulent mixing are both relevant, and the critical
exponents depend on $d$, $\xi$ and the compressibility parameter $\alpha$.

Practical calculations of the critical exponents and the regions of
stability for all the regimes were performed in the one-loop approximation
of the RG, which corresponds to the leading order of the double expansion
in two parameters $\xi$ and $\varepsilon=4-d$. It has shown that, for the
both models, compressibility enhances the role of the nonlinear terms in
the dynamical equations. The region in the $\varepsilon$--$\xi$ plane,
where the new nontrivial regime (4) is stable (the corresponding fixed
point is positive and IR attractive), is getting much wider as the degree
of compressibility increases. In its turn, turbulent transfer becomes more
efficient due to combined effects of the mixing and the nonlinear terms.

Let us illustrate these general statements by the example of a cloud of
particles, randomly walking in a nearly critical turbulent medium. The
mean-square radius $R(t)$ of a cloud of such particles, started from the
origin at zero time, is related to the linear response function
(\ref{respd}) in the time-coordinate representation as follows:
\begin{eqnarray}
R^{2}(t) = \int d{\bf x}\ x^{2}\, G(t,{\bf x}), \quad
G(t,{\bf x}) = \langle \psi (t,{\bf x}) \psi^{\dag} (0,{\bf 0}) \rangle,
\quad x=|{\bf x}|.
\label{Rad}
\end{eqnarray}
For the response function, the scaling relations of the preceding section
give the following IR asymptotic expression:
\begin{eqnarray}
G(t,{\bf x}) = x^{-\Delta_{\psi}-\Delta_{\psi^{\dag}}} \, F
\left(\, \frac{x} { t^ {1/\Delta_{\omega}} }, \,
\frac{\tau}{t^{\Delta_{\tau}/\Delta_{\omega}}}  \right),
\label{Green}
\end{eqnarray}
with some scaling function $F$. Substituting (\ref{Green}) into (\ref{Rad})
gives the desired scaling expression for the radius:
\begin{eqnarray}
R^2(t) = t^{ (d+2 -\Delta_{\psi}-\Delta_{\psi^{\dag}})/\Delta_{\omega} }
\, f \left( \frac{\tau}{t^{\Delta_{\tau}/\Delta_{\omega}}}  \right),
\label{R3}
\end{eqnarray}
where the scaling function $f$ is related to $F$ in (\ref{Green}) as
\[ f (z) = \int d{\bf x}\, x^{2-2\Delta_{\psi}} \, F(x,z). \]
Directly at the critical point (assuming that the function $f$ is finite
at $\tau=0$) one obtains from (\ref{R3}) the power law for the radius:
\begin{eqnarray}
R^2(t) \propto t^\Omega, \quad \Omega \equiv { (d+2
-\Delta_{\psi}-\Delta_{\psi^{\dag}})/ \Delta_{\omega} } =
{(2-\gamma_{\psi}^{*}-\gamma_{\psi^{\dag}}^{*})/\Delta_{\omega} };
\label{R4}
\end{eqnarray}
the last equality following from the relations (\ref{dim}). For the Gaussian
fixed points the usual diffusion law $R(t)\propto t^{1/2}$ is recovered.
For the regimes (3) the exact result $R(t)\propto t^{1/(2-\xi)}$ is derived,
which for the Kolmogorov value $\xi=4/3$ gives $R(t)\propto t^{3/2}$ in
agreement with Richardson's ``4/3 law'' $dR^{2}/dt \propto R^{4/3}$ for a
passively advected scalar impurity. For the other two fixed points the
exponents in (\ref{R3}), (\ref{R4}) are given by infinite series in
$\varepsilon$ (points 2) or $\varepsilon$ and $\xi$ (points 4); the
first-order approximations are readily obtained from the results given
in tables~\ref{tableG} and~\ref{tableA}.

For the case of incompressible fluid ($\alpha=0$), the most realistic values
$d=2$ or~3 and $\xi=4/3$ lie in the region of stability of the passive scalar
regime (3), so that the spreading of the cloud is determined completely
by the turbulent transfer and is described by the power law (\ref{R4})
with the exact exponent $\Omega^{(3)}= 2/(2-\xi)$. As the degree of
compressibility $\alpha$ increases, the boundary between the regions of
stability of the regimes (3) and (4) in the $\varepsilon$--$\xi$ plane
moves such that the region of
stability of the regime (4) is getting wider (see the discussion in
section~\ref{sec:FPS}). When $\alpha$ becomes large enough, these physical
values of $d$ and $\xi$ necessarily fall into the region of stability of the
new regime (4), the advection and the nonlinearity become both important,
and the crossover in the critical behaviour occurs. The new exponent in
(\ref{R4}) can be represented in the form
\begin{eqnarray}
\Omega^{(4)}= \Omega^{(3)} + \delta\Omega, \quad
\delta\Omega = - (\gamma_{\psi}^{*}+\gamma_{\psi^{\dag}}^{*}) / (2-\xi)
\label{R5}
\end{eqnarray}
(we recall that $\Delta_{\omega}=(2-\xi)$ exactly for the both regimes
(3) and (4)). For the Gribov model, substituting the one-loop expressions
for the dimensions $\gamma_{\psi}^{*}$ and $\gamma_{\psi^{\dag}}^{*}$
from the table~\ref{tableG} into (\ref{R5}) gives
\begin{eqnarray}
\delta\Omega = \frac{(3+\alpha)\varepsilon-6\xi}{3(5+2\alpha)(2-\xi)}.
\label{R6}
\end{eqnarray}
Straightforward analysis of the expression (\ref{R6}) shows that, within
the region of stability of the regime (4), the quantity $\delta\Omega$ is
positive and grows monotonically with $\alpha$. Thus the spreading of
the cloud becomes faster in comparison with the pure turbulent transfer,
due to combined effects of the mixing and the nonlinear terms, and
accelerates as the degree of compressibility increases. For the model
{\it A}, the exponents $\Omega$ in (\ref{R4}) for the regimes (3) and (4)
coincide in the one-loop approximation, but the main quantitative
conclusion remains the same: compressibility enhances the role of the
nonlinear terms and leads to the widening of the region of stability of
the full-scale critical regime.

Further investigation should take into account conservation of the order
parameter, its feedback on the dynamics of the advecting velocity field
(mode-mode coupling in the spirit of the model {\it H}\, of critical
dynamics), non-Gaussian character and finite correlation time of the
velocity statistics. This work remains for the future.

\section*{Acknowledgments}
The authors thank L\,Ts Adzhemyan, Michal Hnatich, Juha Honkonen,
Paolo Muratore Ginanneschi and M\,Yu Nalimov for discussions.
The work was supported in part by the Russian Foundation for Fundamental
Research (grant No~08-02-00125a) and the Russian National Program
(grant No~2.1.1/1575).

\end{document}